\documentstyle[epsfig,twocolumn,aps,prl,epsf,floats]{revtex}
\begin{document}
\draft
\flushbottom
\twocolumn[
\hsize\textwidth\columnwidth\hsize\csname @twocolumnfalse\endcsname

%\documentclass[twocolumn]{revtex4}

%\usepackage{graphicx}
%\usepackage{dcolumn}
%\usepackage{amsmath}

%\begin{document}

\title{Electron spectral function and algebraic spin liquid\\
 for the normal state of underdoped high $T_c$ superconductors  }

\author{Walter Rantner and Xiao-Gang Wen}

\address{Dept.\ of Physics, Massachusetts Institute of Technology,
Cambridge, Massachusetts 02139}
\widetext
\date{\today}
\maketitle
\tightenlines
\widetext
\advance\leftskip by 57pt
\advance\rightskip by 57pt

\begin{abstract}

%%%****************** Abstract ******************************************
We propose to describe the spin fluctuations
in the normal state (spin-pseudogap phase) of underdoped high $T_{c}$
cuprates as a manifestation of an algebraic spin liquid.
% We have
%performed calculations within the slave-boson model to support
%our proposal. 
%Under the spin-charge separation 
Within the slave boson implementation of spin-charge separation, the normal state is described by massless Dirac fermions, charged bosons, and a gauge field.
The gauge interaction, as an \emph{exact} marginal
perturbation, drives the mean-field free-spinon
fixed point to a new spin-quantum-fixed-point --
 the algebraic spin liquid.
% where gapless excitations interact at low
%energies.
%The algebraic spin liquid describes the normal state
%of the underdoped high $T_c$ superconductors.
%All correlations in this new
%phase continue to have algebraic decays, but with different exponents than
%the free-spinon fixed point.  
Luttinger-liquid-like line shapes for the electron spectral function are obtained in the normal state and we show how a coherent
quasiparticle
peak appears as spin and charge recombine.

%The electron spectral function in the normal state was found
%to have a Luttinger-liquid-like line shape as observed in experiments.  The
%spectral function obtained in the superconducting state shows how a coherent
%quasiparticle
%peak appears from the incoherent background as spin and charge recombine.
%in the superconducting state.
%%%******************* End of abstract **********************************
\end{abstract}
]
\tightenlines
\narrowtext

{\it Introduction:}
The key property of high $T_c$ superconductors is their Mott insulator
property at half filling. 
After integrating out the excitations above the charge gap at half filling,
the system is described by a generalized $t$-$J$ (GtJ) model
\begin{displaymath}
H = \sum_{(ij)}\big[ J(\vec{S}_{i}\cdot\vec{S}_{j} - \frac{1}{4}n_{i}n_{j})
- t(c^{\dag}_{\alpha i}c_{\alpha j} + h.c.)\big] + ...
\end{displaymath}
which may contain long range and multiple spin couplings indicated by $...$.
Upon doping, the charge carriers form a new non-Fermi-liquid metallic state.
Understanding this new metallic state is the key to understanding high $T_c$
superconductors. 
For underdoped high $T_c$
superconductors, the metallic state has two striking properties. First, the
Fermi surface does not form a closed loop. Second, the electron spectral
function contains no sharp quasiparticle peak. 
Although we cannot derive the above properties from the GtJ
model, we find, in the salve-boson approach\cite{BZA} to the GtJ model,
that a metallic state described by one of the slave-boson mean-field states -- 
the staggered flux (sF) state (which is also called
$d$-wave paired state) -- has a Fermi surface which does not form a closed loop
\cite{WenLee}. The sF state can also explain many other unique properties of
underdoped high $T_c$ superconductors, such as the positive charge and the low 
density of
the charge carriers. Therefore in this paper we will use the sF state as our
starting point to study the electron spectral function in underdoped high
$T_c$ superconductors. 
The effective theory of the sF state is given in Refs. \cite{WenLee,fourauthor},
which contains spinons, holons and a $U(1)$ gauge field as low energy excitations.

The electron spectral function obtained at the mean-field level (ignoring the
$U(1)$ gauge interaction)
\cite{WenLee} has a line shape different from the one measured in
experiments.  In this paper, we include the gauge fluctuations in our
calculation of the electron spectral function. We find that 
the $U(1)$ gauge interaction does not confine the spinons and holons
(at least above a certain energy).
The $U(1)$ gauge interaction turns out to be an exact marginal perturbation
that drives the mean-field spinon fixed point described by free massless Dirac fermions to a new
spin-quantum-fixed-point, which in turn produces a
Luttinger-liquid-like line shape for the spinon spectral function and the
electron spectral function \cite{Anderson}, at least in the very low
doping limit.  We will call this new spin-quantum-fixed-point -- 
the algebraic spin liquid (ASL). 

We also show how the opening of a gap in the gauge field spectrum yields 
spin-charge recombination and
restoration of a coherent peak in the electron spectral function, which has
been observed in the superconducting phase of the cuprates
\cite{Marshall,Norman,Pan,Shen,Ding}.  The mechanism of the gap formation is
as yet not well  understood theoretically.  It can be due to either boson
condensation or confinement caused by
instantons.\cite{FradkinShenker,NagaosaLee} We find that analyzing doping
dependent ARPES results can help to clarify this issue.
If the gap in the gauge field is due to boson condensation
(the Higgs mechanism),
the sharp quasiparticle peak will appear only in the
superconducting phase\cite{Shen}.  
The weight of the sharp quasiparticle peak will
increase as the superfluid density increases,
%(as we lower $T$ or increase the doping $x$),
$Z\propto x (\rho_s)^{2\alpha}$.\cite{Shen,Ding}.  
%, which  qualitatively
%(if not quantitatively) agrees with the experimental observations in
On the other hand, if the gap of the gauge field is
opened via the instanton effect, the weight of the sharp quasiparticle peak
will be proportional to the doping, $Z\propto x$, and the peak may appear
above $T_c$.

{\it Dirac spectrum in high T$_{c}$ superconductors:}
Our experimentally motivated
starting point is the staggered flux state where the
mean-field degrees of freedom are free fermionic spin carrying particles 
(spinons) and charged bosons (holons).
The question of interest to us is whether the mean-field spinons survive the
inclusion of fluctuations, in particular the gauge fluctuations,
around the mean-field state.
In order to analyze this problem we have mapped the lattice effective theory
for the sF state (at zero doping) onto a
continuum theory of massless Dirac spinors coupled to a gauge 
field,\cite{MarstonAffleck}
whose Euclidean action reads
\begin{eqnarray}\label{QED3a}
S&=&\int d^{3}x \sum_\mu \sum_{\sigma=1}^N\bar{\Psi}_{\sigma}
v_{\sigma,\mu} (\partial_{\mu}-ia_{\mu})\gamma_{\mu}\Psi_{\sigma}
\end{eqnarray}
where $v_{\sigma, 0}=1$ and $N=2$,
but in the following we will treat $N$ as an arbitrary integer.
In general $v_{\sigma, 1}\neq v_{\sigma,2}$. However, for
simplicity we will assume $v_{\sigma,i}=1$ here.
The Fermi field $\Psi_{\sigma}$ is a $4\times 1$ spinor which describes
lattice spinons with momenta near $(\pm \pi/2, \pm \pi/2)$.
The $4\times4$ $\gamma_{\mu}$ matrices form a representation of the
Dirac algebra
$\{\gamma_{\mu},\gamma_{\nu}\}=2\delta_{\mu\nu}$ ($\mu,\nu = 0,1,2$)
The dynamics for the $U(1)$ gauge field arises solely due to the screening by
bosons and fermions, both of which carry gauge charge. In the low doping
limit, however, we will only include the screening by the fermion
fields,\cite{KimLee} which yields
\begin{eqnarray}
\cal Z&=&\int Da_{\mu}\exp\Big( -\frac{1}{2}\int\frac{d^3q}{(2\pi)^3}a_{\mu}
(\vec{q})\Pi_{\mu\nu}a_{\nu}(-\vec{q})\Big) \nonumber \\
\Pi_{\mu\nu}&=&\frac{N}{8}\sqrt{\vec{q}^2}\Big(\delta_{\mu\nu}
- \frac{q_{\mu}q_{\nu}}{\vec{q}^{2}}\Big)
\label{Pi}
\end{eqnarray}

{\it Spectral function - Normal state:} We have analyzed the gauge invariant
spinon Green's
function of the above model in a large $N$ expansion. The details of the
calculation will be described elsewhere.\cite{RantnerWen}
Here we just state the result
\begin{equation}\label{FullGreen}
G(\vec{k}) = -iC\frac{k_{\mu}\gamma^{\mu}}{k^{2-2\alpha}} \quad\quad
\alpha = \frac{16}{N}\frac{1}{3\pi^{2}}\Big |_{N=2} =  0.27
\end{equation}
where $C$ is determined by the energy range over which our effective theory
is supposed to be valid.
Note that the above value of $\alpha$ is for the
$v_{\sigma,1}=v_{\sigma,2}=v$
case. $\alpha$ will take a different value if $v_{\sigma,1}\neq
v_{\sigma,2}$.
Comparing this dressed propagator with the free spinon Green's function
$G_{0} = \frac{-ik_{\nu}\gamma^{\nu}}{k^{2}}$ we see that the inclusion of
the gauge fluctuations has destroyed the coherent quasiparticle pole by
changing the exponent of the algebraic decay.
An important result coming out of this calculation is that
the gauge interaction does not generate any mass and/or chemical potential
terms for the spinons. Since the conserved current (that couples to $a_\mu$)
cannot have any anomalous dimension, the gauge
fluctuation represent an exact marginal perturbation whose inclusion at the
mean-field free spinon fixed point yields a new phase with novel algebraic
behavior.
This new quantum fixed point for the spins is the algebraic spin
liquid (ASL) mentioned above.\cite{comm1} 
We see that the ASL state contains no free quasiparticles at low energies.
It is not the confined phase of the $U(1)$ gauge  
field however, which would bind the spinons into a spin wave. The ASL is closer to the deconfined even though there are no {\em free} spinon quasiparticles at low energies. We still say that there is spin-charge separation in the ASL.

We would like to remark that despite many similarities, there is a
difference between our ASL proposal and
the quantum-critical-point (QCP) approach to high $T_c$
superconductors.\cite{qcp}
We do not assume or require a nearby quantum phase transition which gives
rise to a QCP. The ASL can exist as a phase despite the fact that its gapless
excitations interact even at lowest energies.

In the following we will determine the behavior of the physical electron
spectral function from correlations in the ASL.
By virtue of the spin-charge separation implemented in the slave boson
theory, the physical electron operator is a product of a holon and a spinon.
As mentioned above at the mean-field level these two degrees of freedom
propagate as {\em free} particles and in particular since the mean-field boson
condensation temperature $T_{c}\sim 4\pi xt \sim 4000$K (where $t \sim$
400meV and the hole doping concentration $x\sim 0.1$),
we may consider the bosons to be nearly
condensed in the low energy effective theory.
The electron spectral function being a product of charge and spin
propagators is then simply determined through the spinon correlations.
%and
%takes on the form\cite{RantnerWen}
%\begin{eqnarray*}
%G(\vec{q}) &=& \frac{x}{4} \langle
%f_{1}(\vec{q})f^{\dag}_{1}(\vec{q})\rangle
%\nonumber \\
%&&- \frac{x}{4} \langle f_{2}(-\omega,-{\bf q}-{\bf Q})
%f^{\dag}_{2}(-\omega,-{\bf q}-{\bf Q})\rangle \\
%{\bf Q} &=& (\pi,\pi) \quad \vec{q} = (\omega,{\bf q})
%\end{eqnarray*}
%where $x$ is the hole doping concentration arising from the (nearly)
%condensed holons
%and $1,2$ stand for spin up and down respectively.
%The appearance of ${\bf Q} = (\pi,\pi)$ is a remnant of the translational
%symmetry breaking sF mean-field ansatz.
Mapping the continuum fields back onto the lattice fields we can utilize the
result for the dressed spinon propagator in the ASL to see
the effect of the gauge fluctuations on the physical electron propagator.
We find for the electron spectral function\cite{RantnerWen}
\begin{eqnarray}\label{Spectralfunction}
&& A_{+} =
\theta( \omega)\bigg\{  \frac{xC}{4\pi}\sin(\pi\alpha)
\theta(\omega - E_{f})\frac{\omega + \epsilon_{f}}{[\omega^2 -
E^2_{f}]^{1-\alpha}}\bigg\} \\
&& A_{-} =
\theta(-\omega)\bigg\{  \frac{xC}{4\pi}\sin(\pi\alpha)
\theta(-\omega - E_{f})\frac{-\omega - \epsilon_{f}}{[\omega^2 -
E^2_{f}]^{1-\alpha}}\bigg\} \nonumber
\end{eqnarray}
where $ E_{f} \equiv \sqrt{\epsilon_{f}^{2}+\eta_{f}^{2}} $,
$\epsilon_{f}({\bf q}) = -2\tilde{J}\chi(\cos(q_{x}a)+\cos(q_{y}a))$, and
$\eta_{f}({\bf q}) = -2\tilde{J}\Delta (\cos(q_{x}a)-\cos(q_{y}a))$.
$C$ is determined by noting
$\int d\omega d^2q/(2\pi)^2 A_\pm \sim x$.
Even though the momenta in the expressions for the spectral functions run
over all of the Brillouin zone, strictly speaking they should be restricted
to the vicinity of the four Fermi points $(\pm\pi/2,\pm\pi/2)$
where the lattice fermions are well approximated by massless Dirac fermions.

\begin{figure}[tb]
\centerline{ 
\hfil
\epsfysize=35mm
\epsfbox{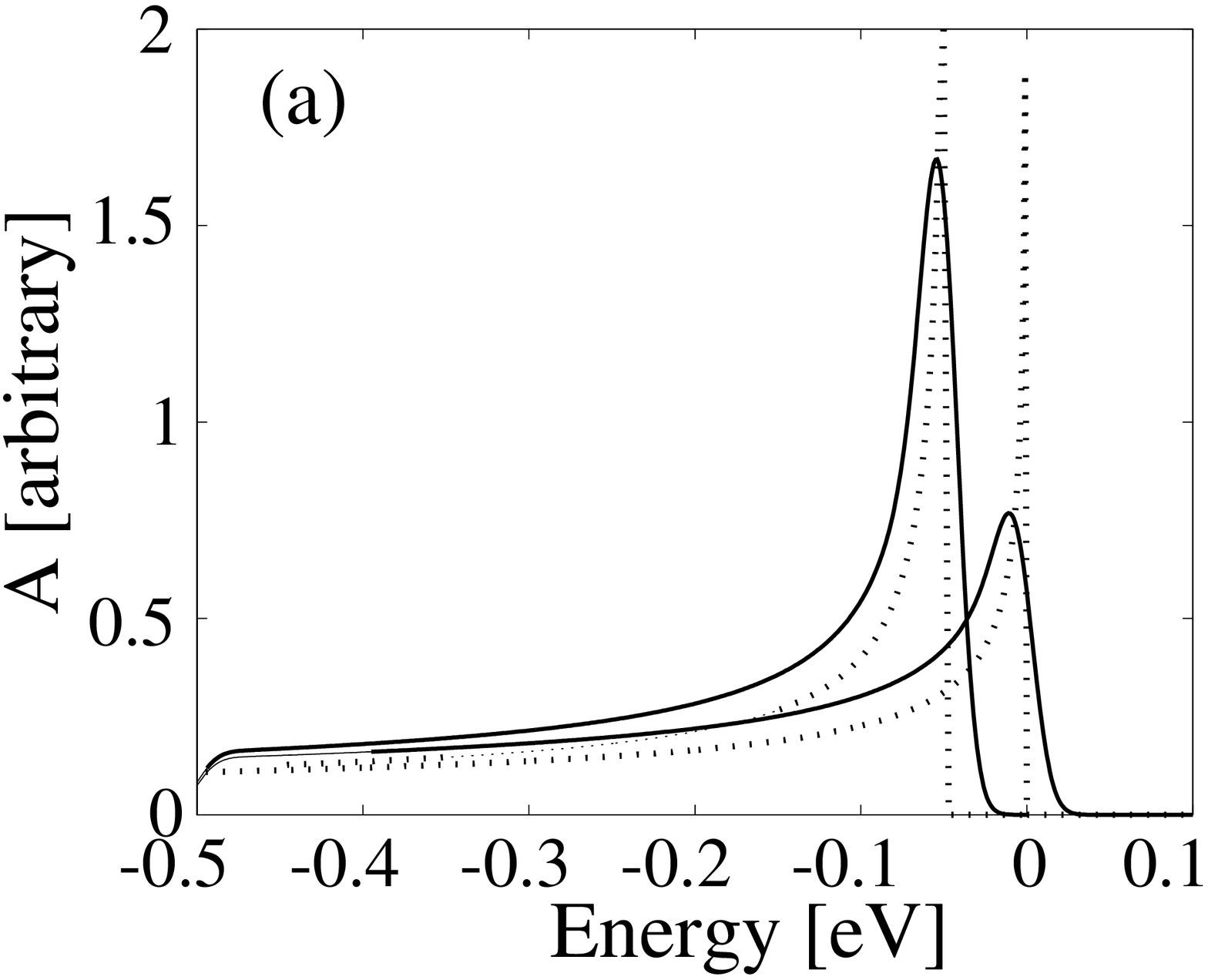} 
\hfil
\epsfysize=35mm
\epsfbox{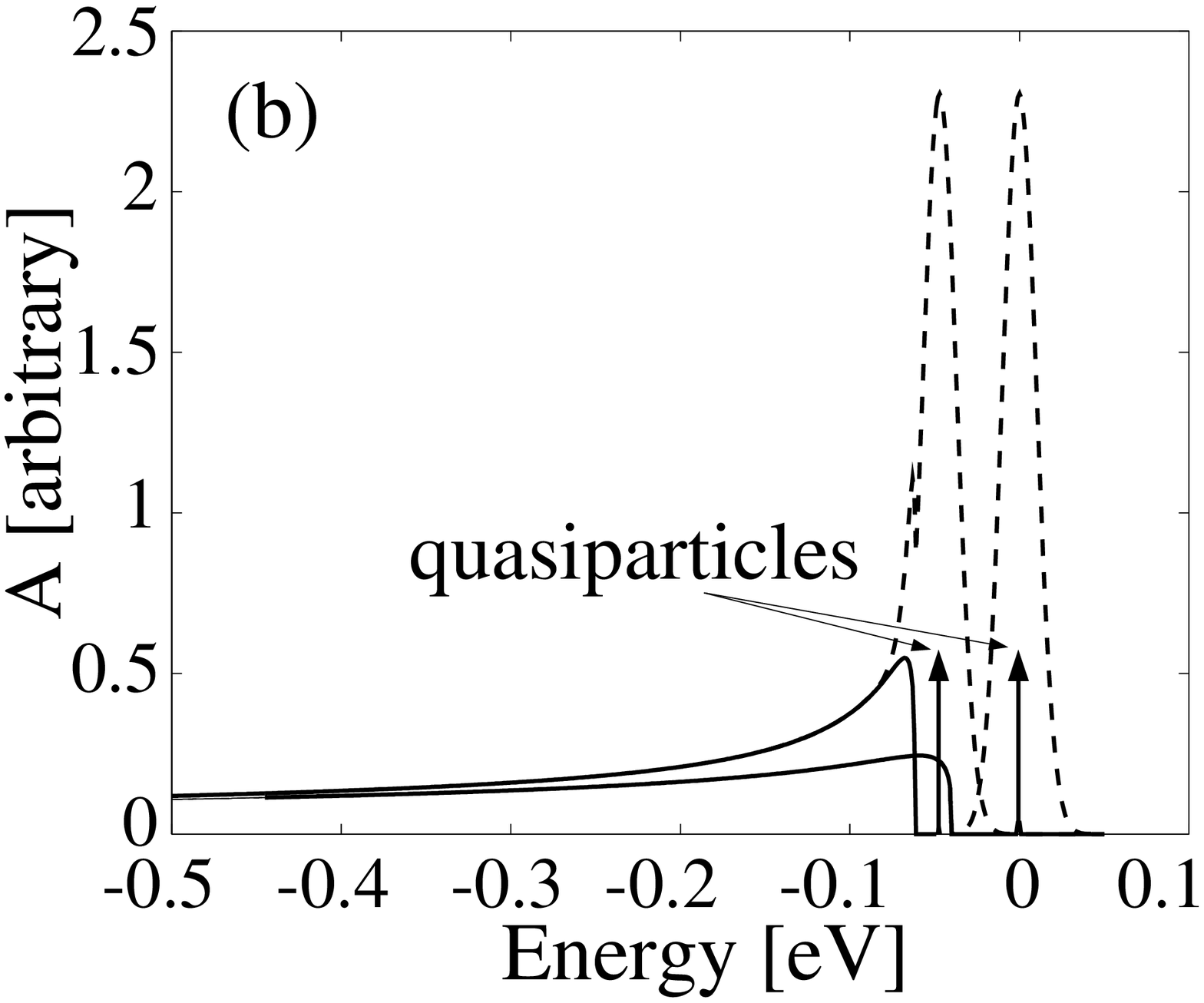} 
\hfil
}
%\begin{center}
%\includegraphics[width=8cm]{Fignormal.eps}
\caption{(a) Two spectra along the $(\pi,\pi)$ direction at $ {\bf q} =
(0.47,0.47)$ and at the node ${\bf q}=(0.5,0.5)$ in units of $\pi/a$. $\alpha
= 0.27$ is used. The
solid line is obtained on smearing the dotted line with a Gaussian of
$\sigma = 10meV$. The important point to note is the lack of a coherent
quasiparticle pole which agrees well with ARPES line-shapes in the normal
state of the cuprates.
(b) Solid line shows spectra at $(0.5,0.5)$ and $(0.47,0.47)$
in the superconducting state with $ m =
40meV$. The dashed line shows the delta function smeared with a Gaussian of
$\sigma = 10meV$ which leads to a break in the line-shape at $(0.47,0.47)$
as opposed to a dip. The arrows indicate the position of the quasiparticle
pole.}
\label{Spectrum}
%\end{center}
\end{figure}

In Fig.(\ref{Spectrum}a) we plot the spectral functions for two momenta
along the zone
diagonal. The main point to note is the lack of coherent quasiparticles in
the spectrum which is in good agreement with experimental results for the
cuprates above the transition temperature.
We would like to stress here again that it is spin-charge separation
combined with the ASL phase for the spin sector that yield
the above spectra without the need for 1D phenomenology.
The incoherent electron spectral function was also obtained using 1D physics
in the stripe model for high $T_c$ superconductor.\cite{kiv}

%Right at the Fermi points $(\pm\pi/2,\pm\pi/2)$,
%the spectral function $A(\omega) \propto
%1/|\omega|^{1-2\alpha}$. For other momenta, the spectral function
%$A(\omega) \propto 1/|\omega|^{1-2\alpha}$ for large $\omega$ and
%%approaches
%$ A(\omega) \propto 1/|\omega-E_f({\bf k})|^{1-\alpha}$
%as $\omega \to E_f({\bf k})$.

{\it Spectral function - Superconducting state:}
In marked contrast to the normal state, the superconducting phase
has been shown to have coherent quasiparticles everywhere in momentum
space below the superconducting gap.
%\cite{Norman,Pan,Shen,Ding}.
%as is evidenced by both ARPES and tunneling measurements
Explaining the development of this coherent behavior
out of the incoherence of the normal state is one of the big challenges in
revealing the high $T_{c}$ physics.  
%After reproducing the normal-state
%electron spectral function employing the spin-charge separation and the
%ASL, we would like to consider the electron spectral
%function in the superconducting state.

In the spin-charge separation picture, the superconducting state can be
obtained through boson condensation in the spin pseudogap phase.
The  gauge field
$a_\mu$ obtains a Higgs mass $m$ which implies that the gauge field is in
the confinement phase.\cite{FradkinShenker}
Thus the spinons and holons are confined in the superconducting phase.
Due to the confinement (which is referred to as
spin-charge recombination) we expect a well defined quasiparticle and
a sharp peak in the electron spectral function to appear in the
superconducting state.
We assume that after gaining a mass gap due to boson condensation (or instanton
effects), the gauge effective theory is described by Eq. \ref{Pi} with
$\Pi_{\mu\nu}=\frac{N}{8}\sqrt{\vec{q}^2+m^2}\Big(\delta_{\mu\nu}
- \frac{q_{\mu}q_{\nu}}{\vec{q}^{2}}\Big)$.
In the boson condensation picture, $m$ is related to the superfluid density
$\rho_s$ (the density of condensed holons):
$\frac{N}{8}m \approx \rho_s/2m_h$, where $m_h$ is the holon mass.
The electron spectral function then takes the form
\begin{eqnarray}\label{spectralmgap}
&& A_{+}(\omega,{\bf q}) =
\Theta(\omega)C\frac{x}{4}\bigg\{(m^2)^\alpha u_{f}^2\delta(E_{f}-\omega)
\nonumber \\
&& + \Theta(\omega^2 - E_{f}^2 - m^2)\frac{
\sin(\pi\alpha)}{\pi} \big[\omega^2 - E_{f}^2 - m^2\big]^{\alpha }
\frac{\omega+\epsilon_{f}}{\omega^2 - E_{f}^2} \bigg\} \nonumber \\
&& A_{-}(\omega,{\bf q}) =
\Theta(-\omega)C\frac{x}{4} \bigg\{(m^2)^\alpha v_{f}^2\delta(E_{f}+\omega)
 \\
&& - \Theta(\omega^2 - E_{f}^2 - m^2)\frac{
\sin(\pi\alpha)}{\pi} \big[\omega^2 - E_{f}^2 - m^2\big]^{\alpha }
\frac{\omega+\epsilon_{f}}{\omega^2 - E_{f}^2} \bigg\}  \nonumber
\end{eqnarray}
where
$v^2_{f} \equiv \frac{E_{f}({\bf q}) - \epsilon_{f}({\bf q})}{2E_{f}} \quad
 u^2_{f} \equiv \frac{E_{f}({\bf q}) + \epsilon_{f}({\bf q})}{2E_{f}}$
are the well known Bogoliubov coherence factors.
In Fig.(\ref{Spectrum}b) we have plotted the spectra for the same momenta 
as in Fig.(\ref{Spectrum}a).
%\begin{figure}[tb]
%\epsfxsize=60mm
%\centerline{ \epsfbox{SCfig1610.eps} }
%%\begin{center}
%%\includegraphics[width=7cm]{Figsuper.eps}
%\caption{Solid line shows spectra at $(0.5,0.5)$ and $(0.47,0.47)$
%in the superconducting state with $ m =
%40meV$. The dashed line shows the delta function smeared with a Gaussian of
%$\sigma = 10meV$ which leads to a break in the line-shape at $(0.47,0.47)$
%as opposed to a dip. The arrows indicate the position of the quasiparticle
%pole.}
%%\end{center}
%\label{Spectralmgap}
%\end{figure}
We can clearly see the two distinct contributions to the spectral function,
the delta function quasiparticle peak and the broad incoherent weight
respectively.
An alternative interpretation of the peak-hump structure was given in
Ref. \cite{Norman1}.
%For comparison with experiment we have singled out the spectral
%function at $(\pi,0)$ in Fig. (\ref{pizero}). The reason for looking at
%$(\pi,0)$ is the higher effective energy resolution achievable by
%%experiments
%at this point. As mentioned before the resolution near the nodes is
%%determined
%by the finite momentum window which in conjunction with the large
%%dispersion
%at the nodes makes detection of the quasiparticle peak difficult.
%\begin{figure}[tb]
%\epsfxsize=80mm
%\centerline{ \epsfbox{Fig3.eps} }
%\begin{center}
%\includegraphics{Fig3.eps}
%\caption{Spectral line at $(\pi,0)$ with a FWHM = 20meV.We chose m=40meV
%%the maximum gap = 30meV and J=130meV. The quasiparticle peak was smeared
%%with a Gaussian to resolution limit.}
%\end{center}
%\label{pizero}
%\end{figure}

%An important question within the gauge field description which we glossed
%over so far is the
As mentioned earlier, there are \emph{different} ways in which the gauge field
acquires its mass. In one picture, the gauge field becomes massive when the
bosons acquire phase coherence via the Higgs mechanism.
%, as we discussed above.
This way the mass generation of the gauge field is tied to the
appearance of the superconducting order.  Even without the boson
condensation, however, the gauge field can acquire a mass via 
instantons,\cite{Poly} which is referred to as the confinement regime.  
In this case the gauge field can be
massive even in the normal state.  We would like to stress that both boson
condensation and instanton effect lead to the same phase where the gauge
field is
gaped and spin and charge recombine.  The two pictures, with different
dynamical
properties, just represent two different limits of the same
phase.\cite{FradkinShenker}
%The above two pictures do not contradict each other. In fact the Higgs
%%phase
%and the confinement phase are actually the same phase.\cite{FradkinShenker}

If the mass comes from boson condensation, then $m$ will be proportional to
the superfluid density. If the mass arises due to instantons, $m$
will be the energy scale below which the instantons become important.
Thus phenomenologically we may put
$ m = m_0+C_1 \rho_s $
to cover both boson condensation and the instanton limit.  In the weak
coupling limit, the mass induced by the instanton, $m_0$, is very small and
the gauge field obtains a noticeable mass $C_1 \rho_s$ only after boson
condensation.  In the strong coupling limit, the gauge field can obtain a
large mass merely through the instanton effect.

In the $SU(2)$ slave boson model, the gauge dynamics and its coupling
constant
is obtained through the screening with the fermions and is of order $1$.  It
is hard to determine from theory if the confinement is caused by boson
condensation or via the instanton effect.
%{}From the experimental fact that
%the sharp quasiparticle peak appears only in (or very close to) the
%superconducting phase,\cite{Shen}
%we see that the weak coupling limit appears to fit the
%experiments better.
%In the weak coupling limit, the gauge mass $m$ is proportional to the
%superfluid density.  
We can see from the expression for the spectral function
(\ref{spectralmgap}) that the separation of the coherent particle peak from
the midpoint of the leading edge of the incoherent background is given by
$\Delta\omega = \sqrt{E^2_{f}({\bf q}) + m^2} - E_{f} $ which simplifies for
the spectrum at the node to $ \Delta\omega = m $. Thus measuring the above
mentioned separation for the spectrum at the node as a function of doping
and
superfluid density via ARPES might give us a clue as to which mechanism is
responsible for the opening of the gap in the gauge spectrum.

{\it Conclusion:} We have shown how the physics of spin-charge separation,
gauge fluctuations, and the algebraic spin liquid
give a consistent way of interpreting the
Luttinger-liquid-like line-shapes seen in the normal state of the cuprates
without resorting to 1D phenomenology.
 We have seen how the gauge fluctuations
destroy the free spinon mean-field phase and drive it to a new fixed point
-- the ASL. On entering the superconducting phase
this ASL is destroyed through the opening of a mass gap in
the gauge fluctuations via either the Higgs mechanism or instantons. This
causes spin-charge recombination.
%Current experimental results favor the Higgs mechanism since the sharp
%quasiparticle peaks only appear below the superconducting transition
%temperature.

We believe that the ASL is a more general phenomenon where gapless
excitations interact even at lowest energy scales. This paper
only discussed a particular realization of the ASL through a slave-boson theory.
It would be interesting to find other realizations of the ASL so that one can
check which one fits experiments better.

%We note that our results on the electron spectral function do not depend on
%the details of the holon sector.
It should be emphasized that in the
spin-charge separation picture adopted in this paper,
the spectral weight in the energy window up to $-4J\sim -0.5$eV
(where $-4J$ is the lower band edge of the mean-field spinons) is mainly
determined by the spinon sector (and the coherent holons)
and is predicted to take the form
$\int_{-4J}^{0}A_{-}(\omega,\vec{q})d\omega d^2q/(2\pi)^2 = a + bx$ 
where $x$ is the doping concentration, $b\sim 1$ and $a \sim$ 0.1.
%Under spin-charge separation
The constant term $a$ arises from the incoherent
holon spectral weight [which is not included in Figs. (\ref{Spectrum}a) and
(\ref{Spectrum}b)].
We can estimate $a$ by noting that the total
mean-field spectral weight for the holons is stretched out from $0$ to
$-8t\sim -3$eV and normalized to $\frac{1}{2}$.\cite{WenLee,fourauthor}
This is important when extracting the doping dependence of the weight of the
quasiparticle peak from ARPES measurements.

In the boson condensation picture, $m \propto \rho_s$ and the weight $Z$ of
the sharp quasiparticle peak can be determined from the superfluid density
$\rho_s$, $Z \propto x (\rho_s)^{2\alpha}$.  {}From this we can determine
the
temperature dependence of the weight of the quasiparticle peak. 
Furthermore, the $T=0$ weight is $Z \propto x^{1+2\alpha}$.
Under the instanton picture, $m\sim m_0$ and we have $Z \propto x$
if $m_0 \gg C_1 \rho_s$.
%In a recent paper D.L. Feng {\it et~al} \cite{Shen} have reported the quasiparticle weight as a function of doping and temperature.
%They note that $Z(T/T_{c})$
%qualitatively resembles the temperature dependence of the superfluid
%%density
%$\rho_s(T/T_{c})$. Following our results we should look at $
%\rho_s^{2\alpha}(T/T_{c}) \sim \rho_s^{1/2}(T/T_{c})$

\begin{figure}[tb]
\epsfxsize=60mm
\centerline{ \epsfbox{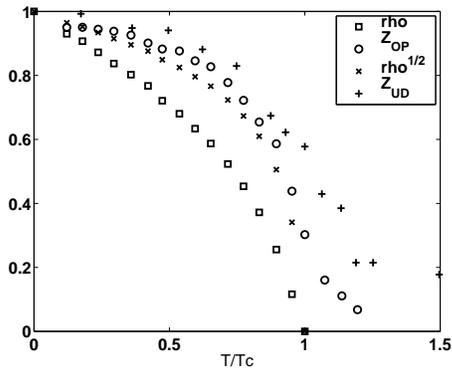} }
%\begin{center}
%\includegraphics[width=8cm]{Fignormal.eps}
\caption{Temperature dependence of $Z_{UD}, Z_{OP}$, $\rho_s$ and $\rho_s^{1/2}$}
\label{SPR}
%\end{center}
\end{figure}

In Fig.(\ref{SPR}) we compare $Z(T/T_{c})$ for underdoped BSCCO (taken from
\cite{Ding}), optimally doped BSCCO (taken from \cite{Shen}) with
$\rho_s^{2\alpha}(T/T_{c}) \sim \rho_s^{1/2}(T/T_{c})$, ($\rho_s(ab)$ for
optimally doped BSCCO taken from \cite{Jacobs}; we wished that $Z$ and $\rho_s$ were
obtained from the same sample.) We observe that $Z$ doesn't
go to zero at $T_c$ and is larger in the underdoped case (with a small T
dependence above $T_c$) which points to mass generation via instantons. Below
$T_c$ we can see how the weight approaches $Z \propto x (\rho_s)^{2\alpha}$
where $x$ is independent of temperature which suggests that the main
contribution to the mass arises through the Higgs mechanism in this
temperature regime.  
%we have plotted the three curves with the curves $\propto Z$ and $\rho_s$
%taken from \cite{Shen}.
%which exhibits nicely $Z \propto x
%(\rho_s)^{2\alpha}$ where $x$ is independent of temperature.

Finally let us note that the behavior of the holons is still poorly
understood. In this paper we have assumed that the holons have a small
energy
scale of order $T_c$ in order to carry out our calculations.
%We have argued
%that the spinons are described by a quantum fixed point -- algebraic spin
%liquid -- in the temperature range $\sim$100K to $\sim$1000K, the holon
%%sector
%may or may not be close to a quantum fixed point in the same temperature
%range.
Although the normal state electron spectral function may not depend on
the details of the holons, many other physical properties, such as normal
state charge transport and the transition to the superconducting state,
require a good understanding of these degrees of freedom.

We would like to thank P.A. Lee, Z.Q. Wang and H. Ding for many very helpful
discussions. W.R. would particularly like to thank A. Abanov for helpful
discussions in the early stages of this project.  This work is
supported by NSF Grant No. DMR--97--14198 and by NSF-MRSEC Grant
No. DMR--98--08941.

Note added in proof: After this paper was submitted a related work appeared
\cite{FranzTesanovich}


\begin{thebibliography}{18}
\bibitem{comm1}
One can show quite rigorously that the ASL can exist as a
stable quantum phase at half filling, even though the gapless
excitations always interact and cannot be described by free fermions at low
energies. The ASL is even stable against the instanton
fluctuations of the gauge field when $N$ is large enough\cite{Poly}
or when $v_{\sigma,1}$
is very different from $v_{\sigma,2}$.\cite{wen1}
\end{thebibliography}
\end{document}